\documentclass[12pt, reqno] {article}
\usepackage{latexsym}
\usepackage{amsmath}
\usepackage{mathrsfs}
\usepackage{amsfonts}
\usepackage{graphicx}

\usepackage[totalheight = 21cm, totalwidth = 17cm]{geometry}
\newcommand{\bea}{\begin{eqnarray}}
\newcommand{\eea}{\end{eqnarray}}
\newcommand{\be}{\begin{equation}}
\newcommand{\ee}{\end{equation}}
\numberwithin{equation}{section}

\def\non{\nonumber}
\def\jac{\vartheta}

\allowdisplaybreaks

\begin{document}
\begin{titlepage}  
\pagestyle{empty}
\baselineskip=21pt
\vspace{2cm}
\rightline{\tt gr-qc/0501060}
\rightline{CERN-PH-TH/2004-243}
\vspace{1cm}
\begin{center}
{\bf {\large Potentials between D-Branes in a 
Supersymmetric Model of Space-Time Foam}}
\end{center}
\begin{center}
\vskip 0.2in
{\bf John~Ellis}$^1$, {\bf Nikolaos E. Mavromatos}$^{2}$ and 
{\bf Michael Westmuckett}$^{2}$
\vskip 0.1in
{\it
$^1${TH Division, Physics Department, CERN, CH-1211 Geneva 23, Switzerland}\\
$^2${Theoretical Physics, Physics Department, 
King's College London, Strand WC2R 2LS, UK}\\
}
\vspace{1cm}
{\bf Abstract}
\end{center}
\baselineskip=18pt \noindent  
%%%%%%%%%%%%%%%%%%%%%%%%%%%%%%%%%%%%%%%%%%%%%%%%%%%%%%%%%%%%%%%%%%%%%

We study a supersymmetric model of space-time foam with two stacks each of
eight D8-branes with equal string tensions, separated by a single bulk
dimension containing D0-brane particles that represent quantum
fluctuations. The ground-state configuration with static D-branes has zero
vacuum energy, but, when they move, the interactions among the D-branes
and D-particles due to the exchanges of strings result in a non-trivial,
positive vacuum energy. We calculate its explicit form in the limits of
small velocities and large or small separations between the D-branes
and/or the D-particles. This non-trivial vacuum energy appears as a
central charge deficit in the non-critical stringy $\sigma$ model
describing perturbative string excitations on a moving D-brane. These
calculations enable us to characterise the ground state of the
D-brane/D-particle system, and provide a framework for discussing brany
inflation and the possibility of residual Dark Energy in the
present-day Universe.

%%%%%%%%%%%%%%%%%%%%%%%%%%%%%%%%%%%%%%%%%%%%%%%%%%%%%%%%%%%%%%%%%%%%%
\vfill
\leftline{CERN-PH-TH/2004-243}
\leftline{December 2004}
\end{titlepage}
\baselineskip=18pt
%%%%%%%%%%%%%%%%%%

\section{Introduction and Summary}

In~\cite{west} we have presented a new D-particle model for supersymmetric
space-time foam in the context of string theory. The model consists of two
stacks of eight D8-branes and their images, each stack being adjacent to
an orientifold plane, which compactifies the bulk space~\cite{ibanez}.
The model has one bulk spatial dimension with a
distribution of  D0-particles. 
When the D-branes and D-particles are
static, a rigorous perturbative string-theory calculation, valid for weak
string couplings $g_s \ll 1$, showed that the vacuum energy of the
configuration is zero, as expected in a supersymmetric string vacuum. This
result was obtained by calculating, in a $\sigma$-model framework for
Dirichlet branes~\cite{polchinski}, the quantum effective potential
between the D-branes and D-particles due to the exchanges of pairs of
open strings between them.

In~\cite{brany}, a scenario for brany inflation was proposed, based on
collisions between branes in this supersymmetric D-brane model for
space-time foam, extending and completing earlier work in this
subject~\cite{gravanis}. A pair of D-branes was assumed to be moving with
a small relative velocity with respect to one another, so that
adiabaticity is valid, and hence also string perturbation theory. The
collision of the D-branes results in a departure from equilibrium, which,
for the perturbative stringy excitations on the observable D-brane world,
is quantified by the dynamical generation of a central charge deficit for
the respective world-sheet $\sigma$ model. From the space-time point of
view, this appears as non-trivial positive vacuum energy. Long after the
collision, when string perturbation theory is applicable on the string
world sheet, the recoil of the branes is described by appropriate
logarithmic conformal field theory deformations, which cause the dark
energy to relax adiabatically. This relaxation is described by the
underlying dynamics of the super-critical (Liouville) string
theory~\cite{ddk,aben} on a recoiling D-brane, in which the Liouville mode
is identified with the target time~\cite{emn}. This identification stems
from the appearance of a surplus of central charge in the model, as
compared to the critical equilibrium value, and is made manifest by the
target-space dynamics, in particular the minimisation of the effective
potential in the appropriate low-energy theory describing string dynamics
on the D-brane~\cite{gravanis}.

An important issue in such a scenario is the initial condition for
inflation, and in general the characterisation of the non-equilibrium
situation following the brane collision. To compute the {\it initial}
central-charge deficit between the colliding branes, very soon after the
collision, lies outside the remit of perturbation theory in general.
However, if, for the sake of simplicity, one assumes an adiabatic
collision as in~\cite{gravanis,brany}, then it is possible to estimate the
initial central-charge deficit by performing standard critical
string/brane theory calculations of annulus amplitudes. These describe the
exchange of open strings when the two branes are close to each other. In
general, for adiabatic motion of the branes, such computations are also
sufficient to provide estimates of the asymptotic value of the
vacuum energy on the observable brane world, when the branes approach
equilibrium long after the initial collision. 

As was shown in~\cite{gravanis}, the form of the central-charge
deficit at times $t$ long after the collision of two branes
recoiling adiabatically with relative velocity $v \ll 1$ (in units of the
speed of light $c=1$), is:
\begin{equation}
Q^2 =C-C^* \sim Q_\infty^2 + \frac{v^4}{t^2}
\label{cdd}
\end{equation}
where $Q_\infty$ is the asymptotic (equilibrium) 
value of the central-charge deficit, and the 
(relaxing) second contribution is the result of the sudden collision of 
the branes,
which is described by appropriate logarithmic terms~\cite{gravanis} in the 
corresponding conformal field theory.
At short times after the collision, 
one can describe the situation for adiabatic collisions with $v 
\ll 1$ by setting $Q^2 = Q_{\it init}^2 (v)$,
where  $Q_{\it init}$ is independent of time.
Both $Q_{\it init}$ and $Q_\infty$ can be determined by 
standard (critical) string/brane theory annulus amplitude computations,
as they can be identified with the potentials that the brane world feels
in the presence of other branes, as a result of the exchange of open string
pairs stretched between the branes~\cite{bachas,douglas,polchinski}. 
The main point of this article is to calculate these potentials explicitly 
in the supersymmetric D-brane model for space-time foam described 
in~\cite{west}. In this model, there are also contributions to the 
effective potential due to the presence of the bulk D-particles,
which are also be taken into account in our discussion below. 

There are various ways in this framework by which branes can collide and
produce inflation~\cite{brany}. According to one scenario, one D-brane
from one stack collides with the other stack, and then bounces back to
collide again with the stack where it originated, where it eventually
stops moving. Another scenario is that two or more D-branes collide, and
later return to their initial positions. In these scenarios, inflation on
the brane world occurs soon after the first collision~\cite{brany}. As
mentioned above, at early times shortly after the collision, the
interaction potential between the colliding D-branes may be expressed as
an (approximately)  constant central charge deficit in the $\sigma$ model
describing string excitations on the brane world. In such a situation, the
(approximately) constant central-charge deficit corresponds to the Hubble
parameter of the inflationary era~\cite{brany}. As the moving D-brane
traverses the bulk space, the recoiling brane world traverses a
non-trivial distribution of D0-particles, which also contribute to the
central charge (Hubble parameter) and affect the D-brane dynamics.

An important point in the analysis of~\cite{brany} was that the asymptotic
state of the configuration, where all the D-branes and D0-particles come
to a standstill, is the supersymmetric vacuum of~\cite{west}, with zero
vacuum energy. However, for finite times, the logarithmic conformal field
theory analysis has demonstrated the existence of a vacuum energy on the
brane that relaxes to zero, scaling with the Robertson-Walker cosmic time
$t$ as $t^{-2}$, as seen in (\ref{cdd}). The vacuum energy is {\it
positive}.  It is crucial in such scenarios to calculate the initial and
final values of the central charge, thereby determining the precise
relations of the Hubble parameter for inflation and the recoil relative
velocities of the branes, as well as the current value of the cosmological
constant in our Universe. These relations are essential if one is to make
contact with the astrophysical data, and thereby constrain the parameters
of the model of~\cite{brany}.

\begin{figure}[tb]
\begin{center}
\includegraphics[width=4cm]{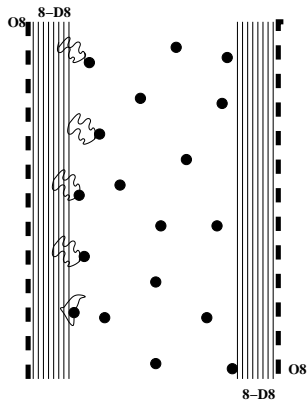}
\end{center}
\caption{\it A model for supersymmetric D-particle foam
consisting of two stacks each of eight parallel coincident D8-branes, 
with orientifold planes (thick dashed lines) attached to them.
The space does not extend beyond the orientifold planes.
The bulk region of ten-dimensional space in which the D8-branes 
are embedded is punctured by D0-particles (dark blobs). 
The two parallel stacks are sufficiently far from each other
that any Casimir contribution to the vacuum energy is negligible.
Open-string interactions between D0-particles and D8-branes
are also depicted (wavy lines). If the D0-particles are stationary or 
moving parallel to the D8-branes,
there is zero vacuum energy on the D8-branes, and the configuration 
is a consistent supersymmetric string vacuum.}
\label{fig:nonchiral}
\end{figure}

The computations presented here apply critical string theory annulus
amplitude calculations to evaluate first the interaction potential between
a D0-particle and a D8-brane or O8 orientifold plane in relative motion.
We then study D8-brane/D8-brane and D8-brane/O8 orientifold plane
potentials. We calculate both long- and short-range limits of the
potentials among the D8-branes and D0-particles, where the range is always
measured relative to the string length $\ell_s = \sqrt{\alpha '}$, where
$\alpha'$ is the Regge slope.  

\begin{figure}[tb]
\begin{center}
\includegraphics[width=4cm]{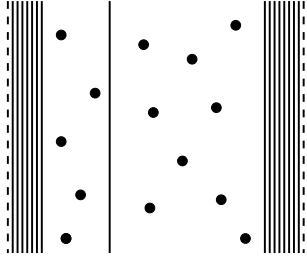}
\end{center}
\caption{\it One of the branes in the stacks of 
Fig.~\protect\ref{fig:nonchiral}
starts moving towards the other stack. As it moves, it crosses D0-particles
in the foam, and the (string) interactions with the 
other branes and D-particles induce a potential on this moving brane.}
\label{seveneight}
\end{figure}

\begin{figure}[tb]
\begin{center}
\includegraphics[width=4cm]{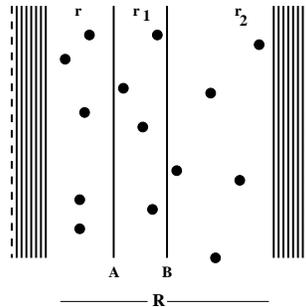}
\end{center}
\caption{\it A scenario with asymmetric colliding branes.
We estimate the effective potential
felt by the D8-brane world A in the environment of the other D8-branes,
D0-particles and O8 orientifold planes.}
\label{fig:asymm}
\end{figure}

These calculations are then applied to the case of a D8-brane moving
between a stack of eight D8-branes and an orientifold on one side, and
seven D8-branes and the other orientifold on the other side, as shown in
in the supersymmetric space-time foam configuration of
Fig.~\ref{seveneight}, as well as scenarios with two moving D8-branes, as
seen in Fig.~\ref{fig:asymm}. As we show, the computations are functions
of even powers of the relative velocity, so that the direction of motion
is irrelevant. These calculations determine the initial values of the
vacuum energy for the different colliding-brane scenarios, and thus the
initial values of the central-charge deficit in the respective $\sigma$
models. These values should be matched smoothly with the relaxing value of
the vacuum energy, long after the collision, and determine the way in
which the vacuum energy approaches zero after reheating. The computations
also shed light on the possible value of the vacuum energy near a second
collision, where the moving D-brane world may eventually stop, in a model 
with of two moving D-branes, cf, Fig.~\ref{fig:asymm}.

We perform the relevant calculations case by case, following the
conventions of~\cite{bachas,polchinski}, examining first the short- and
long-range potentials in various simplified cases, and then
combining the results to obtain the effective interaction felt by the
moving D-brane world in the scenarios of Figs.~\ref{seveneight} and
\ref{fig:asymm}, with their respective stacks of D-branes, orientifolds
and bulk D0-particles.

\section{D0-Brane Potentials}

The potentials  are evaluated by means of annulus world-sheet
calculations, representing a stretched pair of open strings
emanating from a point of one D-brane and ending up on another point
on the other brane. 
As explained in detail in~\cite{douglas}, to leading order
in a weak string coupling $g_s$, to which we 
restrict ourselves throughout this work, 
in the open-string channel the annulus is equivalent to a trace over
all open-string states 
$$V_n^{\text{lightest}} \sim v^n \int d^{p+1}k\int_0^\infty dt 
\hspace{2pt}t^{n-1}e^{-t(k^2+m^2)}~,$$ 
where $p$ is the number of Neumann-Neumann coordinates, 
leading to an expansion 
in powers of $v^2/r^4$, where $r$ is the distance between the
D-branes, for small velocities $v \ll 1$. 
Convergence of this series 
implies the existence of a new characteristic 
minimum length $r_M$, shorter than the 
string length $\ell_s$~\cite{douglas}:
\begin{equation}
r_M \simeq \sqrt{v}\ell_s~,\qquad v \ll 1
\label{shortdistance}
\end{equation}
which we use later in the article, when we 
discuss the short-distance behaviours of the various potentials.
 
In what follows, we calculate the potentials between moving branes in both
short- ($\ell_s \gg r \ge r_M, v \ll 1$) and long-distance
regimes ($r \gg \ell_s$), as compared to the string length $\ell_s$, in
the context of the supersymmetric model of space-time foam described in
\cite{west}. We start with simple cases that pave the way to the final
configuration, which involves moving D8-branes in a bulk filled with
supersymmetric D0-particles, that may themselves be moving~\footnote{It is
a general feature of such calculations that motion of a brane (or a
D0-particle) parallel to the direction of another brane does not lead to
any force between the two. Therefore, we always consider transverse
relative motions.}.

\subsection{D8-Brane/D0-Particle Potential}
\subsubsection{Short Range}
We remind the reader that the annulus amplitude $\mathcal{A}$ is the
integral over target time of the potential between the branes,
$\mathcal{V}$. In what follows, we suppress (for brevity)  volume factors
$V_p$, where $p$ is the number of Neumann coordinates, as well as an
overall proportionality factor of $g_s^2$, where $g_s$ is the (weak)
string coupling, which appears in front of the world-sheet graphs.
These factors will be re-instated at the very end of our computations.

The static D-brane potential between a D0-particle and a D8-brane is
\cite{elementary, west}
\bea
\mathcal{V} &=& -\int \frac{dt}{2t}(8 \pi^2 \alpha^\prime t)^{-1/2} e^{-R^2 t/ 2
  \pi\alpha^\prime} \left[\frac{f_2^8(q)-f_3^8(q)+ f_4^8(q)}{f_4^8
    (q)} \right] \non\\
&=& -\frac{1}{2} T_0 R \left[1 - 1 \right].
\eea
When the D0-particle is moving with velocity $v$ transverse to the
D8-brane, this potential is modified to 
\bea\label{westres2} 
&&\mathcal{V} =  -\int  \frac{dt}{2t}
e^{-R^2t/(2\pi\alpha^\prime)}\non\\&&
\Bigg\{\left(\frac{\jac_3(0,q)}{\jac_1^\prime(0,q)}\right)^{-1} 
  \left(\frac{\jac_2(0,q)}{\jac_4(0,q)}\right)^{4}
\left(\frac{i\jac_3(v t,q)}{\jac_1(v t,q)}\right)_{NS}
-\left(\frac{\jac_2(0,q)}{\jac_1^\prime(0,q)}\right)
^{-1}
  \left(\frac{\jac_3(0,q)}{\jac_4(0,q)}\right)^{4} 
  \left(\frac{i\jac_2(v t,q)}{\jac_1(v t,q)}\right)_{R}\non\\   
&&-\delta_{(\Delta-8){(-1)^F R}} \Bigg\},
\eea
where $\Delta=8$. The $(-1)^F R$ sector is divergent because the
velocity changes the ghost and superghost zero-mode properties. This
divergence is cancelled later on by the potential between the
D0-brane and a O8 orientifold plane. 

To find the  potential ${\cal V}$ between the two branes, 
it is convenient to re-express  the annulus world-sheet diagram ${\cal
  A}$  in terms of a proper time ($\tau$) integral: 
\bea
\mathcal{A} &=& \int_{-\infty}^\infty d\tau \mathcal{V}(v, r^2)\non\\
\mathcal{V}(v,r^2) &=& -\int_0^\infty \frac{dt}{2t} \sqrt{t} v
\frac{e^{-r^2 t/2\pi\alpha^\prime}}{\sqrt{2\pi^2 \alpha^\prime}}
\Bigg\{\dots \Bigg\},
\eea
where the quantity in the \{ \} brackets is the same as 
the corresponding one in (\ref{westres2}), 
$r^2 \equiv R^2 +v^2 \tau^2$, and the additional factors are to
counteract the effect of the integral
\be
\int_{-\infty}^{\infty} d \tau \exp\left[{-t[R^2+v^2
      \tau^2]/2\pi\alpha^\prime}\right] = e^{-t R^2}
\frac{\sqrt{2\pi^2\alpha^\prime
}}{2v\sqrt{t}}[\text{erf}(\infty) - 
  \text{erf}(-\infty)] = e^{-t R^2}
\frac{\sqrt{2\pi^2\alpha^\prime }}{v\sqrt{t}}.
\ee
The short-range  potential corresponds to the case $ r \ll \ell_s$, 
which formally corresponds to taking the 
$t\rightarrow \infty$ limit, keeping  $r^2 t$ finite but arbitrary. 
Notice that the minimum-distance condition (\ref{shortdistance}) 
is mathematically compatible with this limit, assuming
small velocities, $v \ll 1$.

In that case, expanding the factor in curly brackets to the lowest
order in $v$ gives
\be
\mathcal{V}(v,r^2) = -\int_0^\infty \frac{dt}{4t^{3/2}}  v
\frac{e^{-r^2 t/2\pi\alpha^\prime}}{\sqrt{2\pi^2 \alpha^\prime}}
[-1+\frac{1}{3}t^2v^2]
\ee
giving
\be\label{pot1}
\mathcal{V}= \hspace{1pt}-\frac{r}{4\pi\alpha^\prime} - \frac{\pi\alpha^\prime
  v^2}{12r^3}. 
\ee
Some remarks are now in order. We first 
notice that the velocity-independent term is due to 
the fact that an isolated D8-brane is not permitted,
due to flux conservation~\cite{polchinski}. 
As we discuss in the next subsection, this term will indeed be cancelled 
when we consider the case of a D8-brane in the presence of an
orientifold. This is consistent from the point of view of flux
conservation because the overall RR charge in a compact space must be zero.
Secondly, we observe that the minimum 
short-distance condition (\ref{shortdistance}), guarantees that
the second term in (\ref{pot1}) is less than $1/\ell_s$, which renders 
well defined an effective low-energy field 
theory with such a potential on the brane. 

\subsubsection{Long Range}

To find the velocity dependence of the long-range potentials, the
$t\rightarrow 0$ limit must be taken,
by applying the converse arguments to the short-range case, as 
before. The exponential terms in the
Jacobi Theta functions are $q=e^{-\pi t}$ thus a modular transformation
taking $t\rightarrow 1/t$ is required. Properties of the Jacobi Theta
functions, including modular transformations, can be found in the
appendix. Expanding the potentials before after the modular
transformation gives  
\bea
\label{D0-D8-long}
\mathcal{V}_{D0-D8}^{long} = -\int
\frac{d\tau}{4\sqrt{\tau}}\frac{(-1 +v^2/2)}
     {\sqrt{2\pi^2\alpha^\prime}} e^{-r^2
       /2\tau\pi\alpha^\prime} = -\frac{r}{4 \pi\alpha^\prime} + \frac{rv^2}{8\pi\alpha^{\prime}}.
\eea

\subsection{O8-Plane/D0-Particle Potential}
\subsubsection{Long Range} 
The orientifold potential is less straightforward than the
annulus. The velocity dependent amplitude is most easily constructed
in the closed string channel, and then modular 
transformed to the open string channel to recover the short range
behaviour. In the closed string channel,
the orientifold interaction takes the form of a cross-cap, which
complexifies the distance and twists 
the oscillators.

The static open string potential between a D0-particle and an
O8-plane~\cite{elementary} involves the interaction of the D0-particle
with its image, thus the potential is the sum of an annulus and a
M\"obius graph. In the open string channel this is 
\bea
\mathcal{V} &=& -\int \frac{dt}{4t}(8 \pi^2 \alpha^\prime t)^{-1/2}
e^{-4R^2  
t/ 2  \pi\alpha^\prime} \left[\frac{f_3^8(q)-f_4^8(q)- f_2^8(q)}{f_1^8
    (q)} + 16\times  \frac{-f_4^8(iq)+f_3^8(iq) + f_2^8(iq)}{f_2^8
    (iq)}  \right] \non\\
&=& 8 T_0 R \left[1 + 1 \right].
\eea

The difference in dimension between the brane and the orientifold
gives the potential the same form as a $Dp-Dp^\prime$ potential but
with the significant difference that the boundary conditions are no
longer $NN$, $DD$ or $ND$ but twisted versions, thus the theta
functions within are different. 

To obtain the form of the potential for the moving case, it is
simplest to start with the open string annulus interaction and then
move to the closed string channel. From there it is straightforward to
transform the amplitude into a cross-cap (giving long distance
behaviour) and then modular transform
back into the open string channel to obtain the short distance
behaviour. 

The open string annulus oscillators have the form 
\be
\frac{f_3^8(q)-f_4^8(q)- f_2^8(q)}{f_1^8  (q)}.
\ee
where the terms correspond to NS, NS$(-1)^F$ and R respectively.
Substituting $t\rightarrow 1/t$ and modular transforming into the
closed string channel  gives the corresponding cylinder amplitude
\be
\frac{f_3^8(r)-f_2^8(r)- f_ 4^8(r)}{f_1^8  (r)},
\ee
where $r = e^{-\pi\tau}$ and the closed string channel NS-NS 
sector (first two terms) corresponds to the open string NS and
NS$(-1)^F$ terms and the closed string R-R sector corresponds to the
open string R term.

The action of the orientifold on the oscillator trace manifests as a
twist in the oscillators and a modification of the argument of the
functions $r \rightarrow ir$:
\bea
\alpha_n^{0,9} &\longrightarrow& (-1)^n \alpha_n^{0,9} \non\\
\alpha_n^{1,\dots, 8} & \longrightarrow&   -(-1)^n\alpha_n^{1,\dots,8},
\eea
giving\footnote{The M\"obius amplitude has contributions from 
NS$(\Omega I_9)$, NS$((-1)^F\Omega I_9)$ and R$((-1)^F\Omega I_9)$
\cite{elementary}, so, using the argument above, the cylinder R term
changes to R$((-1)^F\Omega 
I_9)$. The $(-1)^F$ projection is combined with the $\Omega I_9$
projection, leaving it unchanged.}
\be
\frac{-f_4^8(ir)+ f_3^8(ir)+f_2^8(ir)}{f_2^8  (ir)}.
\ee
where the last term is the R-R sector contribution $R((-1)^F\Omega
I_9)$, which is divergent in the moving case for the same reasons as
the D8-D0 case. As will be seen below,
when one considers the combination of 16 D8-branes and an O8-plane, an
exact cancellation of terms in the individual potentials occurs,
including the divergences from the O8-D0 and D8-D0 interactions. This
is a demonstration of why Type IA string theory has the form that it
does, requiring two O8-planes and 32 D8-branes for stability. 

When the D0-brane is moving  transverse to the orientifold, the static
case becomes 
\bea \label{closedd0}
&&\mathcal{V}(v,r^2)=  -\frac{2v}{\sqrt{2\pi^2\alpha^\prime}}\int
\frac{d\tau}{\pi \sqrt{\tau} }
e^{-4r^2/(2\tau\pi\alpha^\prime)}
\left(\frac{\jac_1^\prime(0|i\tau+1/2)}
     {\jac_1(v|i\tau+1/2)}\right)\times \non\\&& 
     \Bigg\{\left(\frac{\jac_3(0|i \tau+1/2)}  {\jac_2(0|i
       \tau+1/2)}\right)^{4}   \frac{\jac_4(v i\tau+1/2)}  {\jac_4(0|
       i\tau+1/2)} -  
     \left(\frac{\jac_4(0|i\tau+1/2)} {\jac_2(0|i \tau+1/2)}\right)^{4}
     \frac{\jac_3(v |i\tau+1/2)} {\jac_3(0|i\tau+1/2)} \Bigg\}.
\eea
As the D0-brane and O8-plane have different dimensions, the velocity
dependence of the potential will be $\mathcal{O}(v^2)$, thus when
considering these interactions in isolation the
annulus graph corresponding to the interaction of the D-brane with its
image does not to contribute. Due to the specific arrangement of
branes under consideration, the cancellations mentioned above mean
that this term solely determines the form of the potential. Expanding and
integrating 
(\ref{closedd0}) gives 
\be
\mathcal{V}_{D0-O8}^{\hspace{1pt}long} = \frac{4r}{ \pi\alpha^\prime}
- \frac{2 
  rv^2}{\pi\alpha^\prime}
\ee
Summing the contributions from the 16 D8-branes with that from the orientifold
plane, the overall potential cancels exactly (the cancellation occurs
at the level before the expansion in powers of the velocity). The annulus 
graph coming
from the interaction of the D0-brane with its image is now the only overall
non-zero contribution, giving
\be\label{d0d0pot}
\mathcal{V}_{D0-D0}^{\hspace{1pt}long} = \frac{-15\pi^3\alpha^{\prime
    3}v^4}{2r^7}. 
\ee

\subsubsection{Short Range}

As before, to find the short-range potential, a modular transformation
must be performed. To convert a cross-cap into a M\"obius strip
requires the transformation $\tau = 1/4t$, details of which can be
found in the Appendix. Using this transformation and writing in terms
of a proper-time integral gives
\bea 
&&\mathcal{V}(v,r^2) =  -\frac{4v}{\sqrt{2\pi^2
    \alpha^\prime}}\int  \frac{dt}{\pi\sqrt{t}}
e^{-4r^2t/(2\pi\alpha^\prime)} 
\left(\frac{\jac_1^\prime(0|it+1/2)}
     {i\jac_1(2vt|it+1/2)}\right)\times \non\\&& 
  \Bigg\{\left(\frac{\jac_4(0|it+1/2)}{\jac_2(0|it+1/2)}\right)^{4}
\frac{\jac_3(2vt |it+1/2)}{\jac_3(0|it+1/2)}
-   \left(\frac{\jac_3(0|it+1/2)}{\jac_2(0|it+1/2)}\right)^{4} 
 \frac{\jac_4(2ivt |it+1/2)}{\jac_4(0|it+1/2)} \Bigg\},
\eea
Expanding and integrating, we find
\be
\mathcal{V}_{D0-O8}^{\hspace{1pt}short}=\frac{4r}{\pi\alpha^\prime} -
  \frac{2\pi\alpha^\prime v^2}{3r^3}. 
\ee
Summing the contribution from the 16 D8-branes with this result  shows
that the velocity-independent terms are cancelled:
\be\label{shortcancel}
\mathcal{V}^{\hspace{1pt}short} =
16\left(-\frac{r}{4\pi\alpha^\prime} - \frac{\pi\alpha^\prime 
  v^2}{12r^3}\right) + \frac{4r}{\pi\alpha^\prime} -
  \frac{2\pi\alpha^\prime v^2}{3r^3} = -\frac{2\pi\alpha^\prime
    v^2}{r^3}, 
\ee
in agreement with \cite{danielsson, probe}.

\subsubsection{Discussion}

When there are equal numbers of D8-branes on either side of the 
D0-particle, the terms linear in $r$ are cancelled, leaving an 
overall long-range potential
\be\label{veldep}
\mathcal{V}_{long} =  - \frac{15\pi^3\alpha^{\prime 3}v^4}{r^7}.
\ee
For the dynamical situations under consideration in later sections, the
distribution of  D8-branes is asymmetric, thus this cancellation will
not occur. For an asymmetric case, terms independent of the velocity
would be present in the potentials. 

This would appear to be a problem, since velocity-independent terms
mean that one cannot have a  vacuum configuration. The solution is via
the mechanism of  string creation~\cite{creation,creation2}, which is
related to the fact that between two equal sets of D8-branes the
effective low-energy theory is Type IIA supergravity, whereas between
an asymmetric distribution of branes there is massive Type IIA
supergravity. 

We recall that the system we consider~\cite{west} is Type IA string 
theory~\cite{polchinski},
with the following configurations of branes: one orientifold
plane located at each of the fixed points 
of the $S_1/\mathbb{Z}_2$ orbifold, $X^9 =0$ and $X^9 = R$, and eight
D8-branes and their images sitting on each orientifold. In the bulk there
are some D0-particles, distributed so that interactions between them are
negligible. A moving D0-particle in this situation feels a force $\propto
v^2$ or $v^4$, and hence zero force at zero velocity.

\begin{figure}[tb]
\begin{center}
\includegraphics[width=4cm]{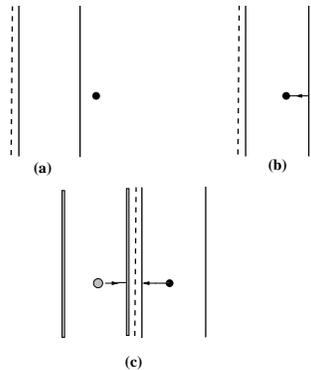}
\end{center}
\caption{\it Single string creation due to a D0-particle traversing a 
D8-brane in the presence of orientifolds, in the model of~\cite{west}. (a) 
The D0-particle (circle) is on the right side of an isolated D8-brane
(thin vertical line), which itself is on the right side of a stack
of D8-branes (thick line) and an orientifold O8-plane (dashed line).
(b) The D0-particle (circle) crosses the D8-brane, creating an 
elementary string, each half of which is attached on one side of the 
brane. (c) The interaction of the D0-particle with its image, as it 
approaches the stack, due to the presence of the orientifold O8-plane. 
Images of D8-branes and D0-particles with respect to the O8-plane are 
denoted by grey shading.}
\label{fig:stringcreation}
\end{figure}

If one of the D8-branes were to move adiabatically across a D0-particle,
there would be an unequal number of D8-branes on either side of the
D0-particle. Naively, using only string perturbation theory, the
potential felt by the D0-particle would then include linear
terms as in (\ref{D0-D8-long}). However, as discussed in~\cite{creation2},
in Type IIA supergravity the D8-brane is the source of a 10-form
field strength, with dual $F = \star F_{10}$ that couples to the
D0-brane as
\be
\mu_0 \int d\tau F A_0.
\ee
Here $F$ is piece-wise constant, which means that when the D0-particle 
traverses the D8-brane it jumps by $\mu_8$, and so $\mu_0 F$ jumps by
$\mu_0 \mu_8 = 1/2\pi\alpha^\prime$. This is interpreted as the
creation of a string between the branes, which cancels the linear
force from the annulus diagram, see Fig.~\ref{fig:stringcreation}. 
As discussed in \cite{elementary}, when a D0-brane is in a region of
massive supergravity with fundamental strings between the D0-brane and
D8-branes, the D0-brane can only move adiabatically in the transverse
direction. Thus D0-branes in between an asymmetric configuration of
D8-branes give no contribution to the vacuum energy; only those in
massless Type IIA backgrounds can move. The string creation mechanism
ensures that the overall energy of the system is velocity dependent
and that there always is a supersymmetric vacuum in the static case.

The maximum distance between the orientifold planes is defined by the
harmonic function in the D8-brane metric. When the distribution of branes
on the orientifold planes is not symmetric the harmonic function blows up
at~\cite{harmonic} $|X^9| = |X^9_{\text{critical}}| = 1/(g_s |m|)$, where
$g_s$ is the string coupling and $|m|$ is the mass parameter associated
with the massive supergravity present in the bulk. When the situation is
symmetric, the bulk supergravity returns to normal Type IIA, thus $|m|$ is
zero and the critical distance goes to infinity. In physical terms,
this gives a critial distance $R <
2\pi\ell_s/(n-8)g_s$, where there are $16-2n$ D8-branes at $X=0$ and $2n$
D8-branes at $X=\pi R$. which for the symmetric case $n = 8$ yields no
upper bound on $R$.

\section{Interactions of a Moving D8-Brane}

We now consider the situation where a D8-brane is moving towards the
stack of 8 D8-branes and the O8-plane. The same approach as for the
D0-O8 interaction is used, with the main difference being that as the
D8-brane and O8-plane have the same dimension the annulus contribution
of $\mathcal{O}(v^4)$ cannot be discarded. 

\subsection{D8-Brane/D8-Brane Interaction}
\subsubsection{Short Distance}
The static potential between two D8-branes vanishes
\bea
\mathcal{V} &=& -\int \frac{dt}{2t}(8 \pi^2 \alpha^\prime t)^{-9/2} e^{-R^2
  t/ 2   \pi\alpha^\prime} \left[\frac{f_3^8(q)-f_2^8(q)-
    f_4^8(q)}{f_1^8 (q)} \right] = 0.
\eea
The potential for the case of a moving brane is similarly
\bea
&&\mathcal{V}  =  -\int  \frac{dt}{2t}(8\pi^2\alpha^\prime t)^{-8/2}
e^{-R^2t/2\pi\alpha^\prime}(2\pi)^3\times\non\\&&
\Bigg\{\left(\frac{\jac_3(0,q)}{\jac_1^\prime(0,q)}\right)^{3} 
\left(\frac{i\jac_3(\nu t,q)}{\jac_1(\nu t,q)}\right)
-\left(\frac{\jac_2(0,q)}{\jac_1^\prime(0,q)}\right)
^{3}   \left(\frac{i\jac_2(\nu t,q)}{\jac_1(\nu t,q)}\right)
-\left(\frac{\jac_4(0,q)}{\jac_1^\prime(0,q)} 
  \right)^{3}   \left(\frac{i\jac_4(\nu t,q)}{\jac_1(\nu
    t,q)}\right) \Bigg\},\non\\
\eea
which expands to
\be
\mathcal{V} = \frac{r v^4 }{2^{13} \pi^9 \alpha^{\prime 5}}
\ee
at short distances and also the same for long distances 
\bea
\label{8-8lr}
\mathcal{V} (v,r^2)_{D8-D8}^{long} &=& -\int_0^\infty \frac{dt}{2t^{3/2}}(8\pi^2
\alpha^\prime)^{-4} \frac{v^4}{\sqrt{2\pi^2 \alpha^\prime}}e^{-r^2
  t/(2\pi\alpha^\prime)} \non\\
&=& \frac{r v^4}{2^{13} \pi^{9} \alpha^{\prime 5}}.
\eea

\subsection{D8-Brane/O8-Plane Interaction}
\subsubsection{Long Range}
The extension of the D8-brane/D8-brane static interaction to the 
D8-brane/O8-plane case is a simple extension of the previous result:
\bea
\mathcal{V} = -\int \frac{dt}{4t}(8 \pi^2 \alpha^\prime t)^{-8/2}
e^{-4R^2 t/ 2 
  \pi\alpha^\prime} \left[\frac{f_3^8(q)-f_4^8(q)- f_2^8(q)}{f_1^8
    (q)} + 16\times  \frac{-f_4^8(iq)+f_3^8(iq)- f_2^8(iq)}{f_1^8
    (iq)}  \right]\non\\ .
\eea
Performing the same procedure as before to find the velocity
dependent orientifold contribution yields
\bea
&&\mathcal{V}(v,r^2) = -\frac{v}{\sqrt{2\pi^2\alpha^\prime}}\int\frac{d\tau}{4\sqrt{\tau}}
e^{-4r^2/2\tau\pi\alpha^\prime} (8 \pi^2 \alpha^\prime)^{-4}(2\pi)^3\times
\non\\ &&
\Bigg\{\left(\frac{\jac_3(0|i\tau)}{\jac_1^\prime(0|i\tau)}\right)^{3}
\left(\frac{\jac_3(v |i\tau)}{\jac_1(v|i\tau)}\right)  
-\left(\frac{\jac_2(0|i\tau)}{\jac_1^\prime(0|i\tau)}\right)^{3} 
\left(\frac{\jac_2(v |i\tau)}{\jac_1(v|i\tau)}\right)
-\left(\frac{\jac_4(0|i\tau)}{\jac_1^\prime(0|i\tau)}\right)^{3}   
\left(\frac{\jac_4(v |i\tau)}{\jac_1(v|i\tau)}\right)
\non\\&&
+16\left(   
\left(\frac{\jac_4(0|i\tau+1/2)}
     {\jac_1^\prime(0|i\tau+1/2)}\right)^{3}
     \left(\frac{\jac_4(v|i\tau+1/2)} {\jac_1(v|i\tau+1/2)}\right)
     -\left(\frac{\jac_3(0|i\tau+1/2)}
     {\jac_1^\prime(0|i\tau+1/2)}\right)^{3} \left(\frac{\jac_3(v
       |i\tau+1/2)}{\jac_1(v|i\tau+1/2)}\right)\right.  \non\\&&
+\left.\left(\frac{\jac_2(0|i\tau+1/2)}
{\jac_1^\prime(0|i\tau+1/2)}\right)^{3}
\left(\frac{\jac_2(v|i\tau+1/2)}{\jac_1(v|i\tau+1/2)}\right)
\right)\Bigg\} 
\eea
Expanding,
\bea
\mathcal{V} (v,r^2)_{D8-O8}^{long} &=&
-\int_0^\infty 
\frac{d\tau}{4\sqrt{\tau}}(8\pi^2 
\alpha^\prime)^{-4} \frac{-15v^4}{\sqrt{2\pi^2
    \alpha^\prime}}e^{-4r^2 
  /2\tau\pi\alpha^\prime} \non\\
&=& \frac{-15rv^4}{2^{13} \pi^{9} \alpha^{\prime 5}},
\eea
the form of the potential reflecting the fact that it is interacting with an object of
similar dimension with negative RR charge.

\subsubsection{Short Range}
Applying the modular transform $\tau \rightarrow 1/4t$, the short
range potential is 
\bea
&&\mathcal{V} =
-\frac{v}{\sqrt{2\pi^2\alpha^\prime}}\int\frac{dt}{4\sqrt{t}}
e^{-4r^2t/2\pi\alpha^\prime} (8 
\pi^2 \alpha^\prime t)^{-4}(2\pi)^3\times
\non\\ &&
\Bigg\{\left(\frac{\jac_3(0|it)}{\jac_1^\prime(0|it)}\right)^{3}
\left(\frac{\jac_3(\nu t|it)}{\jac_1(\nu t|it)}\right)  
-\left(\frac{\jac_4(0|it)}{\jac_1^\prime(0|it)}\right)^{3} 
\left(\frac{\jac_4(\nu t|it)}{\jac_1(\nu t|it)}\right)
-\left(\frac{\jac_2(0|it)}{\jac_1^\prime(0|it)}\right)^{3}   
\left(\frac{\jac_2(\nu t|it)}{\jac_1(\nu t|it)}\right)
\non\\&&
+\frac{16}{8}\left(   
-\left(\frac{\jac_3(0|it+1/2)}{\jac_1^\prime(0|it+1/2)}\right)^{3}   
\left(\frac{\jac_3(2vt|it+1/2)}{\jac_1(2vt|it+1/2)}\right) 
+\left(\frac{\jac_4(0|it+1/2)}{\jac_1^\prime(0|it+1/2)}\right)^{3}
\left(\frac{\jac_4(2v t|it+1/2)}{\jac_1(2v t|it+1/2)}\right)\right.   
\non\\&&
+\left.\left(\frac{\jac_2(0|it+1/2)}{\jac_1^\prime(0|it+1/2)}\right)^{3}
\left(\frac{\jac_2(2v t|it+1/2)}{\jac_1(2v t|it+1/2)}\right)
\right)\Bigg\}
\eea
which upon expansion gives
\be
\mathcal{V}^{\hspace{1pt}short}_{D8-O8} =  - \frac{15rv^4}{2^{13}\pi^9
  \alpha^{\prime 5}}. 
\ee
Summing the contributions from the 16 D8-branes in the stack
and the orientifold contribution gives the result
\be
\mathcal{V}_{16\times D8-D8}+\mathcal{V}_{D8-O8} =
\frac{+rv^4}{2^{13}\pi^9\alpha^{\prime 5}}.
\ee
It is interesting to note that the brane-image interaction has an
important effect in both the long- and short-range cases.

\section{Space-Time Foam Configurations}

In light of the calculations presented above, a number of interesting
situations can be considered, which are linked to the supersymmetric
space-time foam construction of~\cite{west}. This construction includes
two stacks of D8-branes and their images, due to orientifold planes that
are attached to each stack, as well as a gas of D0-particles in the bulk
dimension.

\subsection{Vacuum Configurations} 

We consider Type-IA string theory in the configuration where two
orientifold planes sit on the fixed points of the $S_1/\mathbb{Z}_2$
orbifold, at $X^9 =0$ and $X^9 = R$. Eight D8-branes and their images sit
at $X^9=0$, seven (plus images) at $X^9 = R$. Clearly, the D8-branes can
have no relative motion in the vacuum state. The bulk contains a gas of
D0-particles which is sufficiently dilute that the interactions among the
D0-particles are negligible. However, the D0-particles may, in principle,
be moving. If all are moving parallel to the D8-branes, then, as mentioned
previously, they exert no force on the D8-branes. But, in the case of
intersecting (or bent~\cite{west})  D-brane configurations, which are
required in certain constructions in order to obtain chiral matter
localised at intersections (or foldings), there is no longer a parallel
direction. In this case, there would be a unique vacuum configuration, in
which the D0-particles are completely static~\cite{west}.

\subsection{D8-Branes Moving in a Dilute Gas of D0-Particles}

We now consider a configuration in which one D8-brane has separated itself
from the stack and moves adiabatically into the bulk.

As described above, when the moving D8-brane passes by a D0-particle,
charge conservation requires~\cite{creation} that a string be attached
between the D8-brane and the D0-particle.  When the D8-brane passes by a
D0-particle, the D0-particle enters a region of massive Type-IIA
supergravity, as there are 2$\times$9 D8-branes to the left and 7$\times$2
to the right. From the R-R field version of Gauss' Law, for charge to be
conserved a fundamental string must be created between the D8-brane and
the D0-particle. When the D8-brane moves further away from the
D0-particle, it is energetically favourable for the fundamental string to
be replaced by one stretching between the D0-brane and its image, as seen
in Fig.~\ref{fig:stringcreation}. The potential of the configuration is 
\bea
\mathcal{V}_{Total} &=& 14\times D8(r,v) +16\times D8(R-r,v) + O8(r,v) +
O8(R-r,v) \non\\
&=& \frac{(R-2r)v^4}{2^{13}\pi^9 \alpha^{\prime 5}},
\eea
where $D8$ denotes the interaction between two D8-branes, with relative
velocity $v$, and $O8$ the
interaction between the D8-brane and the orientifold.
We note from the above equation that the potential energy is positive
provided $r < R/2$. This indicates an instability of the
configuration, which has a tendency to relax to its equilibrium
position, in which the moving brane returns to its original stack.

In the model of~\cite{west}, the D8-brane interacts with the gas of
D0-particles at both long and 
short range. As shown before, the long-range D0/D8-O8
potential may be represented as the interaction of the D0-particle with 
its image, and falls
off like $r^{-7}$, and so is negligible at large distances. When a
D8-brane moves into the bulk, the cancellation due to the orientifold
plane still occurs, as the D8-branes centre of mass is still over the
orientifold \cite{danielsson}. Hence the potential felt by the
D0-brane close to the D8-brane is the repulsive force (\ref{shortcancel}):
\be\label{d0d8im}
\mathcal{V}=-\frac{2\pi\alpha^\prime v^2}{r^3}.
\ee
As mentioned at the beginning of Section 2, there is a shortest
effective length scale~\cite{douglas,bachas,polchinski}
(\ref{shortdistance}): $r \approx  \alpha^{\prime
1/2}v^{1/2}$. Using this, the D0-particle potential becomes
\be
\mathcal{V}\approx-\frac{2\pi\alpha^\prime
v^2}{\alpha^{\prime 3/2} v^{3/2}}  \approx
-\left(\frac{v}{\alpha^\prime}\right)^{1/2}.
\ee
The total potential is then estimated to be (re-instating, for
completeness, the eight-brane volume factors $V_8$ where appropriate) 
\be
\label{totalone}
\mathcal{V}_{TOTAL} \approx
-N\left(\frac{v}{\alpha^\prime}\right)^{1/2} +\frac{V_8}{2^{13} \pi^{9}
  \alpha^{\prime 5}}\left[v^4 (R-2r)\right],
\ee
where $N$ denotes the number of D0-particles near the D8-brane, and $r$ 
denotes the distance between the D8-brane and the stack of seven
D8-branes.  

The behaviour of this system is physically in agreement with the
stability of Type IA string theory. In the initial situation, there
are eight D8-branes and their images on top of each orientifold. 
In an orientifold compactification models with
two stacks of $n$ branes each, one obtains the theoretical constraint 
$R < 2\pi\ell_s/(n-8)g_s$, which for our case $n = 8$ yields no upper bound on
$R$. For the case where there are seven branes on one orientifold and
nine on the other, the critical distance is $R< 2\pi\ell_s/g_s \approx
\mathcal{O}(10 \ell_s)$ for weak string coupling. Thus if we start
with a symmetric configuration with large $R$, one cannot move a brane
to the opposite orientifold without decreasing the separation, which
would be unphysical.  

\subsection{Two Moving D8-Branes in the Foam}

We now consider the case when two D8-branes, one from each stack, move
asymmetrically into the bulk with non-relativistic velocities $v_1$,
$v_2$, in a direction transverse to their planes, as seen in
Fig.~\ref{fig:asymm}. In this case, more complicated potentials occur:
\bea
\label{twobrane}
\mathcal{V} &=& 14D8(r,v_1)+14D8(R-r,v_1)+14D8(r_2,v_2)+14D8(R-r_2,v_2)
\non\\ &&+O8(r,v_1)+O8(R-r,v_1)+O8(r_2,v_2)+O8(R-r_2,v_2)
+2D8D8(r_1,v_1+v_2)\non\\  
&=&
\frac{-v_1^4R-v_2^4R+
  2r_1(v_1-v_2)^4}{2^{13}\pi^9   
  \alpha^{\prime  5}}.
\eea
where $R=r+r_1+r_2$. In general, the potential does not have a
definite sign as the branes 
move in the bulk, indicating an instability of the configuration,
which relaxes to equilibrium when the two moving branes return to their
original stacks. In the symmetric case where $v_1=-v_2=v$, we have 
(re-instating eight-brane volume factors)
\be
\label{twosymm}
\mathcal{V}_{sym} = V_8\frac{(30R-64r)v^4}{2^{13}\pi^9\alpha^{\prime 5}},
\ee
which is positive provided $r$ is less than $15R/32$.

In the presence of D0-particles close to the moving
D8-branes, short-distance potentials occur, which 
are the dominant contributions due to the fact that 
long-range D0-particle/D8-brane potentials fall like $r^{-7}$ (\ref{d0d0pot}). 
Using the short-distance substitutions above, an overall contribution of 
order
\be
\label{dparticle}
\sim -N\left(\frac{v}{\alpha^\prime} \right)^{1/2}
\ee
is added to the D8-brane/D8-O8 potential. When the D8-branes move at 
different
velocities, the simplification present in (\ref{twobrane}) does not
occur but the results are qualitively the same.

The total energy of the configuration, therefore,  
is given by summing the results (\ref{twobrane}) and (\ref{dparticle}):
\be\label{total}
\mathcal{V}_{\text{total}}\simeq -N\left(\frac{v}{\alpha^\prime}
\right)^{1/2}+ V_8 \frac{-v_1^4R -v_2^4R+
  2r_1(v_1-v_2)^4}{2^{13}\pi^9 \alpha^{\prime 5}}.
\ee
In the symmetric case which we concentrate on here for simplicity, this
reduces to: 
\be\label{totaltwo}
\mathcal{V}_{\text{total}}\simeq -N\left(\frac{v}{\alpha^\prime}
\right)^{1/2}+ V_8\frac{(30R-64r)v^4}{2^{13}\pi^9\alpha^{\prime 5}}.
\ee
As in the previous subsection, the total 
potential is positive, and thus the configuration is stable,  
for a sufficiently small eight-density of 
D0-particles near the D8-brane.

\section{Physical Applications: Inflation and Dark Energy on the Brane 
World}

The calculation in the previous Section gave the potential energy of the
configuration of the space-time foam model~\cite{west} in the case of
moving D8-branes.  As mentioned in the Introduction, such computations
provide information on the energy density that an observer on the
(physical) brane world would detect. For string excitations which are
confined on this brane, this would imply that their perturbative dynamics
is described by a non-critical string $\sigma$ model, with a 
central-charge surplus computed from the above potential energies.

In particular, in the colliding-brane scenario for inflation presented
in~\cite{brany}, the total energy of the configuration is given by
(\ref{totalone}) in the scenario when only one brane moves, collides with
the other stack and bounces back. Alternatively, when two branes from
opposite stacks collide and bounce back, one would have (\ref{total}) (or
(\ref{totaltwo}) in the symmetric case), to which we restrict ourselves
from now on for simplicity. For an observer on the physical brane world,
taken for definiteness to be the brane A in Fig.~\ref{fig:asymm}, the
expressions (\ref{totalone}) (or (\ref{totaltwo})) lead to bulk
contributions to the effective Dark Energy observed on the brane.

The total energy measured by an 
observer on the brane located at 
a position $X^9 = r$ (see Fig.~\ref{fig:asymm}) is:
\be
{\cal V}_{\rm brane} = \int_0^{R} dX^9 V^{(8)}\delta (X^9 - r) + 
\int_0^R dX^9 \rho (X^9,v),
\ee
where $V^{(8)}$ is the D8-brane tension~\footnote{This arises from 
possible quantum corrections
of strings on the D8-brane, and may be assumed to vanish, when a 
sufficient number of supersymmetries exist on the brane.}, 
and $\rho$ is the bulk energy density, such that: 
\be
\int_0^R dX^9 \rho (x^9,v)= \mathcal{V}_{\rm total} 
\ee
with $\mathcal{V}_{\rm TOTAL}$ given by (\ref{totalone}) 
(or (\ref{totaltwo})) in the symmetric two-brane case), 
in the situation in which the energy is positive.

For a $\sigma$ model describing both open-string excitations on the brane
world and closed-string excitations propagating in the bulk, the quantity
$\rho(x^9,v)/V_8$, where $V_8$ is the eight-volume, defines the
central-charge surplus $Q^2= C -C^*>0$, as measured with reference to the
critical value $C^*$.  In our case (\ref{totalone}) (or (\ref{totaltwo}))
this is positive, under the conditions specified previously,
and Liouville dressing~\cite{ddk} is necessary to
restore world-sheet conformal invariance.  The resulting string theory on
the moving brane world is then supercritical~\cite{aben}, so we can apply
the considerations of~\cite{emn,brany,gravanis} and identify the Liouville
zero mode with the target time. In fact, following the precise scenario
of~\cite{gravanis,brany} we may apply target-space dynamical arguments,
supporting energetically this identification.

The central charge surplus, then, is nothing other than the 
vacuum energy density 
of the effective low-energy ten-dimensional bulk field theory, which is 
given by: 
\be
\label{centralcharge}
Q^2 = \rho/V_8~.
\ee
To calculate $\rho (X^9,v)$, we first re-write $N=n V_8$, 
where $n$ is the 
eight-dimensional density of defects near 
the D8-brane world, and then use (\ref{d0d8im}) for the D0-particle
contributions to the total bulk energy in (\ref{totalone}) 
or (\ref{totaltwo}). Taking these into account, we may write,
cf, (\ref{d0d8im}):
\be\label{final}
\rho (X^9, v) = V_8 \left(\frac{c_1 v^4}{2^{13}\pi^9\alpha^{\prime5}} 
- \Theta(r-X^9) \frac{c_2 v^4}{2^{13}\pi^9
\alpha^{\prime5}} + n\delta_{r_\ell}\frac{6\pi\alpha^\prime v^2}{r^4} \right),
\ee
where $c_1 = 1 (30) $ and $c_2= 2 (64)$ in the case 
(\ref{totalone}) ((\ref{totaltwo})), and $\Theta(r-X^9)$ is the Heaviside step function, 
meaning that the result is
non-zero  (and unity) only for
$0 \le X^9 \le r$ (see Fig.~\ref{fig:asymm}), and 
$\delta_{r_\ell}$ is non zero and unity only when $r + r_\ell \ge X^9 \ge r -
r_\ell$, where $r_\ell$ is the characteristic short-distance
scale (\ref{shortdistance}). 

We notice that in the limit of very small velocities, $v \to 0$, 
such that the $v^4$ term is not dominant, it is the D0-brane 
contributions that make the leading-order contribution to the
central-charge deficit. For consistency of the Liouville 
conformal field theory~\cite{ddk}, one should insist that
$Q^2$ never changes sign, as the theory flows to a fixed point on the 
world-sheet renormalization-group trajectory. In our case, the
model theory is supercritical: $Q^2 > 0$~\cite{ddk,aben}.

We now remark that, in the cosmological model of \cite{brany}, inflation
occurs relatively soon after the initial collision, which corresponds to
the case considered in (\ref{final}), with the stack of branes far away
from the moving D8-brane world. It was assumed in~\cite{brany} that the
bulk D0-particle gas is sufficiently dilute during the inflationary era
for the dominant contributions to (\ref{final}) to come from the
D8-brane interactions. We also remark that, in the model of
\cite{brany}, compactification of the extra dimensions on the branes is
necessary, in order to arrive at physically realistic situations. This is
a separate delicate issue, especially because of the presence of the
orientifolds. Moreover, chiral matter is usually achieved in such 
scenarios
by intersecting brane configurations, with chiral string matter localised
on the intersection. We leave these detailed model issues to future works.

Within the uncertainties in numerical factors associated with
these important model details,
the inflationary era is then described by 
(\ref{final}).
It is not unreasonable to assume that, for adiabatic motion of the 
D-branes
during inflation, the density of D0-branes near the moving D8-brane
world adjusts itself so that the central charge surplus 
of the corresponding $\sigma$ model is approximately constant, 
dominated by the first term on the right-hand-side of (\ref{final}),
i.e.,
\be
\label{ccdinit}
Q^2_{init} \sim 4 \cdot 10^{-9}~c_1~v^4~g_s^2.
\ee
where $c_1 = 1 (30) $ for the single (two) moving brane scenario, and 
$g_s$ is the weak string coupling constant, which we have re-instated
here for completeness. 

In the inflationary model of~\cite{brany}, 
one considers an effective gravitational field theory 
on a brane world with a four-dimensional space-time,
in which the effective central charge will acquire
an appropriate compactification volume factor $V_5 R \sim R$.
For this, we assume an orientifold compactification of one large bulk 
dimension 
of size $R$,
and five small, i.e., of size $\ell_s$, compactified D8-brane 
dimensions. 
In such a scenario, 
$QR^{1/2}/3$ (expressed in 
string units with $\ell_s=1/M_s$)
was identified in~\cite{brany},
with the Hubble parameter $H_I$ during inflation, assumed to be constant.
From WMAP data~\cite{wmap} one obtains at the 2-$\sigma$ level:
\be
H_I = \frac{QR^{1/2}}{3} \le 1.48 \times 10^{-5}~M_P,
\label{wmapdata}
\ee
where $M_P \sim 10^{19}$ GeV is the four-dimensional Planck mass, which
is in general  different from the bulk string 
scale $M_s = 1/\ell_s$.

From (\ref{ccdinit}) this implies an upper bound on the 
relative velocity of the moving branes in the symmetric case (\ref{twosymm}):
\be
\label{boundvel2}
v <  0.8 \cdot c_1^{-1/4} \cdot R^{-1/4}\cdot \sqrt{M_P\ell_s/g_s}.
\ee
In realistic string theories,
and also in the models of D0-particle foam we consider 
here~\cite{west,brany} in which there is one large bulk dimension of size 
$R$ which undergoes orientifold compactification, 
and five small, i.e., of size $\ell_s$, compactified D8-brane 
dimensions,  
one may assume $M_P\ell_s \sim 2\sqrt{2}~g_s^{-1}~R^{1/2}$,
yielding 
\be
\label{boundvel}
v \le  1.4 \cdot c_1^{-1/4} \cdot g_s^{-1}.  
\ee
For weakly-coupled strings with $g_s^2 \sim 1/2$, 
which is a value commonly considered as it leads
to acceptable grand-unification-scale gauge couplings in phenomenological 
(supersymmetric) effective low-energy field theories derived from strings, 
we have $ v \le 1.98 \cdot c_1^{-1/4}$, 
indicating that this value of string coupling
is incompatible with the single moving-brane model, $c_1 = 1$.
On the other hand, $g_s^2 \sim 0.5 $ leads to velocities
which are not very non-relativistic, namely
$ v \sim 0.8$ for the model of two branes moving relative to the 
orientifold planes ($c_1 = 30$). Our approximations in the previous 
Sections assumed $v$ to be small enough that $\sin v \sim v$. The above 
value of $v$ satisfies this relation to within 10\%, which may be 
acceptable. One gets compatibility 
for larger values of the string coupling (still less than one), but then
the validity of the weak coupling is in jeopardy.
Moreover, in the colliding-branes scenario of~\cite{brany},
there is the following relation between 
the spectral index $n_S$ for scalar
perturbations and the number of e-foldings $N$: 
\be
\label{ns}
n_S - 1 = -\frac{3}{N}.
\ee
WMAP data~\cite{wmap} yield $n_s - 1 \simeq -4 \cdot 10^{-2}$, which,
on account of (\ref{ns}), implies $N \simeq 75$.

Assuming adiabatic motion of 
branes during inflation, which lasts for a period
of $t_I \sim x\ell_s/v$, with $R > x > 1$ characterising the 
brane separation,
we have $H_I \frac{x\ell_s}{v} = N \simeq 75 $, from which
we obtain, using (\ref{boundvel}), 
a large 
separation of the branes at the end of inflation
for the case of the two moving branes~\cite{brany}:
\be
\label{rmin}
R > x > 2.54 \cdot 10^{6} \cdot c_1^{-1/4} \cdot R^{-1/2}~.
\ee
This implies that $R > 1.06 \cdot 10^4 $.  We do not discuss cosmological
inflation further here, intending to return in a future
publication~\cite{EMNW} to possible scenarios for the end of inflation and
reheating.

Before closing, we remark that the relaxation phenomena discussed
in~\cite{brany}, which result from the recoil of the brane world after the
collision, may be continuing into the present era, in which case they
could contribute to the present-day cosmological Dark Energy. The
relaxation of the recoil-induced vacuum energy density on our brane world
would then receive contributions~\cite{gravanis,brany,EMNDE} of the form
$\rho \sim v^4/t^2$, where $t \sim 10^{60}~t_{\rm Planck}$: $t_{\rm
Planck} = 10^{-43}$~sec is the present cosmic time, and $v$ is the recoil
velocity of the brane just after the collision, which may be of order $v
\simeq 0.8$ in our model.

We are far from claiming a detailed understanding of inflation and Dark
Energy in the above framework. Nevertheless, we believe that our work
provides useful steps towards a consistent mathematical formulation of
inflationary scenarios in the context of non-equilibrium (non-critical)
effective string theories on excited brane worlds. The use of non-critical
strings is highly appropriate for the study of the non-equilibrium
phenomena that dominate Early Universe physics, and may also control the
current evolution of the Universe.

\section*{Acknowledgements}

N.E.M. wishes to thank Juan Fuster and IFIC-University of Valencia 
(Spain) for their interest and support, 
and P. Sodano and INFN-Perugia (Italy) for the 
hospitality and support during the last stages 
of this work. M.W. thanks CERN, Physics
Department (Theory) for their hospitality.
The work of M.W. is supported by an EPSRC (U.K.) Research Studentship.

\section{Appendix - Properties of Jacobi Theta Functions}
\subsection*{Definitions}
\bea
 &&\jac_1(v|it)\equiv\jac_{11}(v|it) =-2 q^{\frac{1}{4}}\sin [\pi v]
\prod_{n=1}^\infty (1-q^{2n}) (1-e^{2i\pi v}q^{2n})
(1-e^{-2i\pi v}q^{2n}) 
\\
&&\jac_2(v|it)\equiv\jac_{10}(v|it) =2
q^{\frac{1}{4}}\cos [\pi v] \prod_{n=1}^\infty (1-q^{2n})
(1+e^{2i\pi v }q^{2n})(1+e^{-2i\pi v}q^{2n}) 
\\
&&\jac_3(v|it)\equiv\jac_{00}(v|it) =\prod_{n=1}^\infty
(1-q^{2n})(1+e^{2i\pi v}{q^{2n-1}}) (1+e^{-2i\pi v}q^{2n-1})
\\
&&\jac_4(v|it)\equiv\jac_{01}(v|it) =\prod_{n=1}^\infty
(1-q^{2n})(1-e^{2i\pi v}q^{2n-1}) (1-e^{-2i\pi v}q^{2n-1})
\eea
where $q=e^{-\pi t}$. The sum representation is also useful
\bea
\jac \left[
\begin{array}{ll}
a \\  b
\end{array}\right]
(v |t )
=\sum_{n= -\infty}^{\infty} \exp\left( 2 \pi i \left[\tfrac{1}{2}
  (n+\tfrac{a}{2})^2t+(n+\tfrac{a}{2})(v+\tfrac{b}{2} \right]\right) 
\eea
as well as  the identity
\bea
\jac \left[
\begin{array}{ll}
a \\ b
\end{array}\right]
\left(v+\frac{\epsilon_1}{2}t+\frac{\epsilon_2}{2}|t\right)
=e^{-\frac{i\pi t\epsilon_1^2}{4}}
e^{-\frac{i\pi \epsilon_1}{2}(2v+b)} e^{-\frac{i\pi
\epsilon_1\epsilon_2}{2}}\jac \left[ \begin{array}{ll}
a +\epsilon_1 \\
b +\epsilon_2 \end{array}\right](v|t) .
\eea

\subsection*{Modular Transformations}
\subsubsection*{Annulus}
\bea
\eta(\tau) &=& (-i\tau)^{-1/2}\eta\left(-\frac{1}{\tau}\right)\non\\
\jac_1 (\nu|\tau) &=& -(-i\tau)^{-1/2}e^{-\pi i
  \nu^2/\tau} \jac_1
\left(\frac{\nu}{\tau}|-\frac{1}{\tau}\right)\non\\
\jac_2 (\nu|\tau) &=& (-i\tau)^{-1/2}e^{-\pi i
  \nu^2/\tau} \jac_4
\left(\frac{\nu}{\tau}|-\frac{1}{\tau}\right)\\
\jac_3 (\nu|\tau) &=& (-i\tau)^{-1/2}e^{-\pi i 
  \nu^2/\tau} \jac_3
\left(\frac{\nu}{\tau}|-\frac{1}{\tau}\right)\non\\
\jac_4 (\nu|\tau) &=& (-i\tau)^{-1/2}e^{-\pi i 
  \nu^2/\tau} \jac_2
\left(\frac{\nu}{\tau}|-\frac{1}{\tau}\right)\non
\eea
\subsubsection*{Orientifold}
The orientifold modular transformations can be easily derived by
following the Appendix of \cite{DiVecchia}.
\bea
\jac_1^\prime (0|i/4t+1/2) &=& (2it)^{3/2}\jac_1^\prime
\left(0|it+1/2\right)\non\\
\jac_1 (v|i/4t+1/2) &=& (2it)^{1/2}e^{-4\pi v^2t } \jac_1
\left(2ivt|it+1/2\right)\non\\
\jac_2 (v|i/4t+1/2) &=& i(2it)^{1/2}e^{-4\pi v^2t } \jac_2
\left(2ivt|it+1/2\right)\\
\jac_3 (v|i/4t+1/2) &=& e^{-i\pi/2}(2it)^{1/2}e^{-4\pi v^2t } \jac_4
\left(2ivt|it+1/2\right)\non\\
\jac_4 (v|i/4t+1/2) &=& e^{i\pi}(2it)^{1/2}e^{-4\pi v^2t } \jac_3
\left(2ivt|it+1/2\right)\non
\eea

\end{document}